\documentclass[a4paper,11pt]{article}
\pdfoutput=1

\usepackage[utf8]{inputenc}
\usepackage{uniinput}
\usepackage{jheppubnew}
\usepackage{bm}

\renewcommand{\(}{\left(}
\renewcommand{\)}{\right)}
\newcommand{\be}{\begin{equation}}
\newcommand{\ee}{\end{equation}}

\newcommand{\eg}{{\itshape e.g. }}
\newcommand{\ie}{{\itshape i.e. }}
\renewcommand{\d}{\mathrm{d}}

\title{de Sitter cosmology on an expanding bubble}
\author{Souvik Banerjee, Ulf Danielsson, Giuseppe Dibitetto, Suvendu Giri, and Marjorie Schillo}
\affiliation{Department of Physics and Astronomy, Uppsala University, Box 516, SE-75120, Uppsala, Sweden}

\emailAdd{souvik.banerjee@physics.uu.se, ulf.danielsson@physics.uu.se, giuseppe.dibitetto@physics.uu.se, suvendu.giri@physics.uu.se, marjorie.schillo@physics.uu.se}

\begin{document}

\abstract{Constructing an explicit compactification yielding a  metastable de Sitter (dS) vacuum in a UV consistent  string theory is an incredibly difficult open problem. 
            Motivated by this issue, as well as the conjecture that all non-supersymmetric AdS vacua must decay, we discuss the alternative possibility of realizing an effective four-dimensional dS cosmology on a codimension-one bubble wall separating two AdS$_5$ vacua.
            The construction further elaborates on the scenario of \href{https://arxiv.org/abs/1807.01570}{\texttt{arXiv:1807.01570}}, where the aforementioned cosmology arises due to a non-perturbative decay and is embedded in a five-dimensional bulk in a time-dependent way.
            In this paper we discuss the relation between this scenario and the weak gravity conjecture and further develop the details of the four-dimensional cosmology.  
            We provide a bulk interpretation for the dS temperature as the Unruh temperature experienced by an accelerated observer riding the bubble.  
            A source of four-dimensional matter  arises from a string cloud in the bulk, and we examine the consequences for the particle mass spectrum. 
            Furthermore, we show how effective four-dimensional Einstein gravity on the bubble is obtained from  the five-dimensional Gauss equation.
            We conclude by outlining some implications that this paradigm will have for holography, inflation, the standard model, and black holes.
}

\maketitle

\section{Introduction}
Over the past decade, evidence has accumulated indicating that it is surprisingly difficult to construct meta-stable dS-solutions in string theory (see \cite{Danielsson:2018ztv} for a review). This has lead to the formulation of the de Sitter (dS) swampland conjecture \cite{Obied:2018sgi, Garg:2018reu, Ooguri:2018wrx}, suggesting that no such dS vacua exists.

In a recent work \cite{Banerjee:2018qey}, we pointed towards a possible new route out of the swampland. Rather than assuming that a string theory realization of dS is a ten-dimensional meta-stable vacuum, we embed an effective dS in a higher dimensional string vacuum in a crucially time-dependent construction. Our starting point is a non-perturbatively unstable, but long-lived, anti-de Sitter (AdS) vacuum that decays to a lower, possibly supersymmetric (SUSY), AdS through bubble nucleation. Solving the Israel junctions conditions across the bubble wall -- taking gravitational backreaction into account -- we find that the induced metric obeys the Friedmann equations with a positive cosmological constant. Furthermore, massive particles are described by the four-dimensional image of strings stretching outwards from the expanding bubble.

This construction is similar to, but crucially different from, the braneworld scenario of Randall-Sundrum (RS) \cite{Randall:1999ee,Randall:1999vf} . The RS construction joins the inside\footnote{Here inside (outside) refers to decreasing (increasing) volume of radial slices as one moves away from the brane.} of two AdS spacetimes across a brane with a $\mathbb{Z}_2$ orbifold. In our construction, the decay of AdS through bubble nucleation necessarily implies that the brane joins two different AdS vacua with an inside and an outside.  The lack of the $\mathbb{Z}_2$ symmetry and the presence of an inside and an outside will lead to different physics on the brane. A positive cosmological constant
is automatic; the condition that the bubble forms through an instanton with finite Euclidean action guarantees that the induced cosmological constant on the brane is positive. 

As stated above, some of the main motivation for this construction is the long-standing difficulty of finding fully explicit dS constructions in string theory and the resulting dS swampland conjectures.  However, the scenario in \cite{Banerjee:2018qey} also draws crucial motivation from another corner of the swampland: the conjecture that all non-supersymmetric AdS vacua must decay \cite{Ooguri:2016pdq,Freivogel:2016qwc} as a consequence of the weak gravity conjecture (WGC) \cite{ArkaniHamed:2006dz}.  By exploring the consequences of this incarnation of the WGC, which has accumulated strong evidence including several recent proofs \cite{Cheung:2018cwt,Montero:2018fns}, we find a loophole in the much more controversial dS swampland conjecture.  Although our construction does not involve a fundamental dS vacuum in string theory, the effective dS on the brane would be indistinguishable from a true dS vacuum to a low-energy observer.  A toy-model exhibiting the presence of the necessary decay channel in a string vacuum of type IIB was presented in \cite{Banerjee:2018qey}, and a fully explicit string embedding remains a promising direction for future work.

Here we will further develop the low-energy cosmological implications of this model. Since our interest is cosmology, we will specialize to the decay of AdS$_5$ giving rise to a $(3+1)$-dimensional cosmology on the brane, however all results generalize to higher dimensions in a straightforward way. 
There have been various attempts at constructing dS cosmologies on branes. Among others, dS cosmologies from brane bending effects have been discussed in \cite{Kaloper:1999sm} and dilatonic braneworld models with self-tuning mechanisms for the four-dimensional cosmological constant have been discussed in  \cite{Amariti:2019vfv,Kiritsis:2019wyk,Ghosh:2018fbx}.
In this paper, we show that the temperature associated with the Rindler horizon of a uniformly accelerating observer in AdS$_5$ is equal to the temperature associated to the cosmological horizon of the four-dimensional dS, thus providing a bulk interpretation of the dS temperature. 
Furthermore, by projecting the five-dimensional curvature onto the expanding brane using the Gauss equation, we derive the four-dimensional Einstein equations on the brane sourced by both matter and gravity.

The organization of this paper is as follows.  In Section \ref{BTWGC}, we demonstrate the equivalence of a form of the WGC applied to higher form gauge fields and the condition that AdS supported by flux can decay through charged membrane nucleation \cite{Brown:1988kg}.  After a brief review of the basic formalism in Section \ref{review}, we discuss the possibility of multiple bubble collisions and the lifetime of a typical dS cosmology.  Then, in Section \ref{rindler}, we explain how the cosmological horizon can be understood as a Rindler horizon in the time-dependent five-dimensional setting. Section \ref{matter} investigates how radiation and matter can be added to our dS cosmology.  We find that four-dimensional matter comes from stretched strings in five-dimensions and express the mass of a particle in terms of the string tension. In Section \ref{4dEinstein}, we generalize the homogeneous and isotropic treatment of the Israel junction conditions using the Gauss-equation. In this way we verify that the induced, low-energy, four-dimensional gravity is described by the Einstein equations. We conclude, in Section \ref{outlook}, with some remarks and outlook regarding applications to holography, speculations on obtaining the standard model and phenomenologically viable inflation, and the inclusion of black holes in this model.

\section{The decay of AdS} \label{BTWGC}
The basis for \cite{Banerjee:2018qey} is the nucleation of a spherical bubble mediating the decay of a non-supersymmetric AdS vacuum to an AdS vacuum with lower energy.  It is not important that the vacuum that nucleates is supersymmetric, and it may be unstable. However, if the conjectures \cite{Ooguri:2016pdq,Freivogel:2016qwc} are true, the process will eventually terminate with the nucleation of a stable supersymmetric AdS.\footnote{Another interesting possibility is the nucleation of a bubble of nothing \cite{Witten:1981gj} such as \cite{Horowitz:2007pr,Ooguri:2017njy}.  It is an interesting question whether a braneworld scenario can exist on the junction between something and nothing.}  Regardless of the stability of the nucleating vacuum, because the bubble is a spacelike surface that asymptotes to the lightcone, subsequent decays on the interior of a bubble will never collide or interact with the original domain wall and are therefore irrelevant to our construction.

Since the abundance of non-supersymmetric vacua in flux compactifications of string theory was first pointed out \cite{Denef:2004ze,Denef:2004cf}, the stability of such vacua has been thoroughly explored (\eg \cite{Danielsson:2016mtx,Danielsson:2017max} in the context of the WGC). In this section we attempt to make explicit the implicit argument behind \cite{Ooguri:2016pdq,Freivogel:2016qwc} which conclude that the WGC applied to higher form gauge fields implies that non-supersymmetric AdS must decay.  We will show that an extension of the higher form WGC \cite{Heidenreich:2015nta} to codimension-one branes is equivalent to the charge-to-tension relationship required for AdS (supported by flux) to be unstable to the nucleation of charged $p$-branes first derived by Brown and Teitelboim \cite{Brown:1988kg,Coleman:1980aw}. We take this equivalence as preliminary evidence in support of the extension of the WGC. Taking these arguments to heart, it would be interesting to revisit the conclusions of \cite{Johnson:2008vn,Narayan:2010em,Danielsson:2016rmq} which find obstacles to certain decay channels involving the thin-wall limit of moduli domain walls in supergravity.  These supergravity domain walls can be forbidden whereas instantons consisting of fundamental branes may still provide non-perturbative decay channels.

The Brown-Teitelboim instanton describes the nucleation of a spherical, charged, codimension-one membrane in a vacuum whose cosmological constant is sourced, at least in part, by a top-form flux (\emph{i.e.} a $d$-form field strength in $d$-dimensions.)  The membrane nucleates and discharges a unit of flux, thereby lowering the cosmological constant.  The instanton solution describes an energy-conserving configuration where the sphere nucleates at rest at a finite size.  The radius of the instanton is determined by the condition that the energy freed-up in reducing the cosmological constant exactly balances the energy cost of the tension of the membrane.  Furthermore, \cite{Brown:1988kg} shows that charged bubbles cannot nucleate in AdS for any arbitrary value of the membrane charge-to-tension ratio.  The reason for this is that if the tension becomes large relative to the charge, the critical bubble will have to grow so that the energy liberated in the volume of the bubble is enough to `pay' for the energy cost of nucleating the membrane.  However, in AdS at large radial coordinate, the volume and the area of a sphere grow with the same power of radius.  Therefore making the bubble larger no longer balances the energy budget.  

The condition for being able to nucleate a bubble arises from the condition that the radius of the critical bubble is real (eqn 5.12g of \cite{Brown:1988kg}):
\be \label{criticalR}
R_*=\left[{2\tilde{\Lambda}_+ \over (d-1)(d-2)} + \sigma_p^{-2} \({1\over d-1}\(Q_p E_+ - \frac12 Q_p^2\) - {M_{d}^{2-d} \sigma_p^2 \over 2(d-1)}\)^2\right]^{-1/2}\,.
\ee
Here, $d$ is the spacetime dimension, $M_{d}$ is the $d$-dimensional reduced Planck mass, $ \sigma_p$($Q_p$) is the tension(charge) of the $p$-brane, $\tilde{\Lambda}_+$ is the cosmological constant exterior to the bubble,\footnote{We will use a tilde to denote the cosmological constant with mass dimension two, with the relation $\Lambda_d = \tilde{\Lambda}_d M_d^{d-2}$.}  and $E_+$ is the `electric field' associated to the top-form flux ($F^{\mu_1\ldots \mu_d}\propto E_+ \varepsilon^{\mu_1\ldots \mu_d}$).  Note that this holds for membrane nucleation with codimension-one, therefore in this result $p=d-2$.

We will consider the case that the cosmological constant outside the bubble is sourced entirely by flux: $\tilde{\Lambda}_+ = -\frac12 M_{d}^{2-d} E_+^2$.  Furthermore, we assume that there are a large number of units of flux: $E_+=Q_p N$ with $N\gg1$.  This is the limit in which higher curvature corrections can be neglected in supergravity.  Solving the condition that the radius of nucleation is positive yields the inequality:
\be \label{tauBT}
\sigma_{d-2}^2 \le {d-2 \over d-1}Q_{d-2}^2 M_{d}^{d-2} -{(d-2)(2d-3) \over (d-1)^2} Q_{d-2}^2 M_{d}^{d-2} {1\over N} + \mathcal{O}\({1\over N^2}\)\,.
\ee
Here we have included the sub-leading correction in $1/N$ to illustrate that it is negative, lowering the tension relative to the charge, and therefore will not provide a loophole for violating the WGC.

It is worth commenting that there is a discrepancy between the assumption $\tilde{\Lambda}_+ = -\frac12 M_{d}^{2-d} E_+^2$ and the Brown-Teitelboim treatment: in \cite{Brown:1988kg} it is assumed that flux always contributes as a positive cosmological constant.  At first glance this seems like a reasonable assumption, \eg, increasing the electric field always increases the energy stored in a capacitor.  However, as we know from Freund-Rubin compactification and many subsequent years of working with flux compactifications, this is not the case.  The most familiar and relevant example is given by the AdS$_5\times S^5$ constructions of type IIB, where the scaling of the cosmological constant with $N$ is given by $\tilde{\Lambda}_5 \propto -N^{-4/3} M_5^2 \propto -N^{-1/2}m_s^2$ (where $m_s$ is the string scale). To reconcile this with the apparent scaling $\tilde{\Lambda}_+ \propto -N^2 M_5^2$ given above, one can check that if the charge, $Q_p$, is taken to scale with $N^{-5/3}$, as it does in the AdS$_5\times S^5$ solution, then the two results agree.

Meanwhile, the WGC for higher form fields \cite{Heidenreich:2015nta}, can be written as,
\be
Q_p^2 M_{d}^{d-2(p+2)} \ge {(p+1)(d-p-3) \over d-2} \({\tau_p \over M_{d}^{(p+1)}}\)^2\,.
\ee
We see that this does not trivially apply to the case of interest, $d=5$, $p=3$, because the right hand side becomes negative.  This does not invalidate the inequality, but it makes it into a trivial statement.  We propose a non-trivial extension of this result, effectively replacing $(d-p-3)\to|d-p-3|$ for the case $p=d-2$. This results in precise agreement with \eqref{tauBT} in the $N\to \infty$ limit.  While this extension is not derived from the extremal limit of a codimension-one black brane, we see that it shares the property that saturation of the inequality corresponds to an extremal, flat, domain wall.  This flat domain wall does not have a finite Euclidean action and therefore cannot nucleate in a decay process, defining the boundary of stability.  These similarities with the WGC applied to higher codimension objects, together with the equivalence to the Brown-Teitelboim result suggests an explicit argument behind the reasoning of \cite{Ooguri:2016pdq, Freivogel:2016qwc}.

\section{Cosmology on a bubble} \label{review}
We begin here with a review of the basic setup in a theory that has two AdS$_5$ vacua: one with a cosmological constant $\tilde{\Lambda}_+ = -6k_+² \equiv -6/L_+²$ and another with a lower vacuum energy: $\tilde{\Lambda}_- = -6k_-² \equiv -6/L_-²$, so that $k_->k_+$. The vacuum with higher energy can then decay through the nucleation of a spherical Brown-Teitelboim instanton.  Unlike the RS braneworld scenario, the spacetimes across this bubble are not related by a $\mathbb{Z}_2$ symmetry \ie the two sides are truly the inside and the outside of a bubble as required for the nucleation of the instanton.  

In global coordinates, the metric inside and outside the bubble is given by
    \begin{equation}\label{gglobal}
    \begin{split}
    \d s² _\pm &= -f_\pm(r)\d t² + f_\pm(r)^{-1}\d r² + r² \d \Omega_3², \\
    \textrm{where}\quad f_\pm(r)&=1 + k²_\pm r² .
    \end{split}
    \end{equation}
    Parametrizing the radius of the brane in terms of proper time for an observer at rest on the bubble as $r=a(τ)$, the induced metric takes the FLRW form,
    \begin{equation} \label{gind}
    \d s²\big|_\mathrm{ind}=-\d τ² + a(τ)²\d \Omega_3².
    \end{equation}
    Discontinuity of the metric across the brane forces the presence of a non-zero stress tensor on the brane, $S_{ab}$, which is given by Israel's second junction condition
    \begin{equation}\label{eq:israel}
        S_{ab} = -\frac{1}{8π G_d}\left(\left[K_{ab}\right]-\left[K\right]h_{ab}\right),
    \end{equation}
    where $G_d$ is the $d$-dimensional Newton constant with $1/(8\pi G_d)=M_d^{d-2}$, and $\left[\cdot\right]$ represents the difference of the corresponding quantity across the brane. $K_{ab}$ is the extrinsic curvature defined as $K_{ab}= n_{α;β}e^α_a e^β_b$ where $n_α$ is a unit vector normal to the brane defined in the direction of increasing transverse volume, and $e^α_a = ∂x^α/∂y^a$ are tangent vectors with $x^α$ labeling bulk coordinates and $y^a$ label coordinates on the brane).
    
Assuming only spherical symmetry of the instanton \eqref{gind} and a brane of constant tension, $\sigma$, \eqref{eq:israel} gives a first order equation for the evolution of the radius, $a(\tau)$:
    \begin{equation}
        σ=\frac{3}{8πG_5}\left(\sqrt{k^2_{-}+\frac{1+\dot{a}^2}{a^2}}-\sqrt{k_{+}^2+\frac{1+\dot{a}^2}{a^2}}\right),
     \end{equation}
    where the dot represents derivative with respect to proper time on the brane.   In the limit that all four-dimensional energy scales (set by curvature and the Hubble scale) are small compared to five-dimensional scales (\ie $1/a\ll k, \dot{a}/a \ll k$), the brane appears as a spatially flat Minkowski space with a critical tension:
    \begin{equation}
    σ_\mathrm{crit} = \frac{3}{8πG_5}(k_- - k_+).
    \end{equation}
    For a brane with a slightly sub-critical tension, $σ=σ_\mathrm{crit}(1-\epsilon)$, with $\epsilon>0$, expanding in $\epsilon$ gives the Friedmann equation 
    \begin{equation} \label{dSFriedmann}
   H^2= \frac{\dot{a}²}{a²}=-\frac{1}{a²}+\frac{8π}{3}G₄Λ₄+\mathcal{O}(\epsilon ²),
    \end{equation}
    with the identifications
    \begin{equation}\label{eq:G4L4}
        G₄= 2\left(\frac{k_+k_-}{k_- - k_+}\right)G_5,\qquad Λ₄= σ_\mathrm{crit}-σ.
    \end{equation}
    This represents a dS universe with positive spatial curvature.  It is also important to note that the corrections to \eqref{dSFriedmann} are independent of $a(\tau)$.  Therefore, neglecting to tune the brane tension to be near-critical will still result in a dS expansion, however the cosmological constant will be a more complicated expression.
    
Examining the size of the cosmological constant relative to the four-dimensional reduced Planck mass we find:
\be
{\Lambda_4 \over M_4^4} = 256π²G_5^2\sigma_{\rm crit} \epsilon \( { k_-k_+ \over k_- -k_+}\)^2 \approx \epsilon k^3 M_5^{-3}.
\ee
In the last equality, and in the remainder of this work, we find it useful to assume no large hierarchy between the interior and exterior AdS radii, leading to the approximation $k_-\sim k_+ \sim k_--k_+\sim k$. Thus, we see that as long as the five-dimensional spacetime is weakly curved ($k <M_5$) the four-dimensional cosmology will be as well.  Furthermore, we are interested in the case of a phenomenologically relevant four-dimensional cosmology, \ie, one with a cosmological constant problem, however we would like the fundamental AdS$_5$ vacuum to be natural.  In this case, a modest hierarchy $k<M_5$ implies a tuning of the brane tension to be very nearly critical with $\epsilon \sim 10^{-120}$.

Finally, it is worth noting that exactly this tuning, where the four-dimensional cosmology presents the observed cosmological constant problem but the five-dimensional vacuum has no extreme hierarchies, guarantees that the lifetime of a typical bubble before collision with another such bubble is longer that the four-dimensional Hubble time.  We will consider collisions in a metastable AdS$_+$ vacuum with initial conditions defined such that at some time, $t=0$ the entire space is in the false vacuum.  Then, not only is a bubble guaranteed to nucleate as long as the decay rate is finite, but due to the infinite spatial volume of the $t=0$ surface, an infinite number of decays will occur.  Choosing `our' bubble to be at the center of AdS, this means that at the same time, another bubble will nucleate near the boundary of AdS.  Because the bubbles undergo constant proper acceleration, asymptoting to a lightcone, they will collide in a time of order the AdS radius: $t_{\rm col}\sim L_+$, measured in the global time coordinate \eqref{gglobal}.

We would like to compare this to the Hubble time, given in a four-dimensional observer's proper time: $\tau_{\rm H}=H^{-1}$. The relation between these two coordinates can be found by insisting that a point on the brane follows a timelike trajectory and results in:
\be
{\d t \over \d\tau} = {\sqrt{f(r) +(\partial_\tau r)^2} \over f(r)}.
\ee
Neglecting spatial curvature so that $H$ is constant and $r(\tau) = r_0{\rm e}^{H\tau}$, one readily finds global time as a function of proper time. Again, neglecting the decaying spatial curvature, we have $H\approx \sqrt{\Lambda_4}/M_4 \approx \sqrt{\epsilon} k$.  Since the bubble nucleates at rest, we furthermore have $r_0 = H^{-1} =\tau_H$.  Then, converting to Hubble time in global coordinates we find: $t_{\rm H} = \left({\rm e} -1\right)/ {\rm e}k + O(\epsilon)$.  Therefore,
\be
{t_{\rm H} \over t_{\rm col}} \approx {{\rm e} -1 \over {\rm e}} \approx 0.6.
\ee
This indicates that we the expect braneworld cosmology to remain relatively uneventful for longer than the age of the universe, but one cannot wait an arbitrarily long time before a collision. 

It is important to note that this result depends strongly on our choice of boundary conditions.  Without the assumption that the entire space is in the false vacuum at $t=0$, one would find that the lifetime of any point before collision is zero due to the infinite volume of the past lightcone \cite{Harlow:2010az}.\footnote{The work \cite{Harlow:2010az} discusses these points in great detail, including implications for holography.  Related lifetimes for dS or general FLRW spacetimes are also discussed in \cite{Kashyap:2015lva}. We thank Ashoke Sen for brining these subtleties and references to our attention.}  The boundary conditions considered here correspond to a cutoff in time, before which the decay of the false vacuum is `turned off.'  One could alternatively replace this cutoff in time with a spatial cutoff, such as a cutoff brane at large radial coordinate similar to the `two-brane' RS scenario.  A spatial cutoff also presents the opportunity for tuning the lifetime before collision relative to the Hubble rate, exponentially increasing the expected time until collision.  Therefore, the possibility of constructing a finite volume AdS may be more relevant for model building.   On the other hand, brane collisions need not spell doom for braneworld observers and may give rise to interesting phenomenology related to inflation and reheating.  We will come back to this point in Section \ref{outlook}.
       
\section{Bulk acceleration and induced temperature} \label{rindler}

The new understanding of how to obtain a positive cosmological constant provided by this model also accounts for the presence of the cosmological horizon. One way to view the cosmological horizon is through its associated temperature. Just as black holes have a Hawking temperature, it is generally understood that there is a temperature -- the dS temperature -- associated with the cosmological horizon \cite{Gibbons:1977mu}. We will now show that this temperature can be understood as the Unruh temperature measured by an accelerated observer.\footnote{The  temperature associated with an accelerating brane has also been discussed in \cite{Russo:2008gb}.}

An observer living on the surface of the expanding bubble accelerates in the radial direction. The temperature measured by an observer in AdS with a proper acceleration $a^\mu$ is given by \cite{Deser:1997ri}
    \begin{equation} \label{TAdS}
    T = \frac{1}{2π}\sqrt{a^μ a_μ - k²},
    \end{equation}
    where $μ$ runs over the five-dimensional bulk.  
    The proper acceleration of an observer in AdS is given by $a^μ = \ddot{x}^μ + Γ^μ_{νσ}\dot{x}^ν\dot{x}^σ$. In global coordinates, \ie \eqref{gglobal},
    \begin{equation}
    a^t = \ddot{t} + \frac{f^\prime (r)}{f(r)}\dot{r}\dot{t}\quad
    \textrm{and}\quad
    a^r = \ddot{r} - \frac{f^\prime (r)}{2f(r)}\dot{r}² + \frac{1}{2}f^\prime(r)f(r)\dot{t}²,
    \end{equation}
    where a prime denotes derivative with respect to the argument and a dot denotes derivative with respect to proper time ($τ$). For a timelike trajectory on the brane, \ie, $-f(r) \dot{t}² + f(r)^{-1}\dot{r}²=-1$, neglecting curvature and parametrizing the radius of the bubble in terms of the Hubble parameter, $r \sim \exp(Hτ)$, the temperature is given by
    \begin{equation}
    \begin{split}
    	T=\frac{1}{2π}\sqrt{\frac{\left(f'(r)/2+\ddot{r}\right)²}{f(r)+\dot{r}²}-k²} =\frac{H}{2π} +{\mathcal O}({\rm e}^{-2H\tau}).
    \end{split}
    \end{equation} 
    The corrections here are consistent with neglecting spatial curvature in a late-time expansion.  From the perspective of the four-dimensional observer, this is the temperature associated with the presence of a cosmological horizon.
Alternatively, one can derive the dS temperature of \cite{Gibbons:1977mu} by considering a four-dimensional dS space with horizon at $r=H^{-1}$ 
    \begin{equation}
    \d s² = -(1-r²H²)\d t² + \frac{1}{1-r²H²}\d r² + r² \d \Omega_2².
    \end{equation}
    The temperature of dS can be computed by performing an analytic continuation of $t\to i t_{\rm E}$ and demanding that there are no conical singularities. This gives $T_\textrm{dS}=H/2π$ which matches the temperature observed by the observer on the expanding bubble at leading order.

\section{Matter fields on the bubble} \label{matter}
Soon after the introduction of the RS scenario it was realized that placing matter in the AdS$_5$ spacetime results in a radiation component in the four-dimensional cosmology \cite{Kraus:1999it}.
 The presence of five-dimensional matter in the bulk results in a Schwarzschild-AdS solution where the function $f(r)$ in equation \eqref{gglobal} becomes $f_\pm(r)=1+r²k²_\pm-8G₅M_\pm/(3πr²)$,\footnote{The coefficient of the $1/r²$ term is chosen such that $M$ is the ADM mass.} 
    Einstein's equations on the brane take the form of Friedmann equations with a spatial curvature, cosmological constant and radiation density:
    \begin{equation}\label{eq:friedmann}
    \frac{\dot{a}²}{a²}=-\frac{1}{a²}+\frac{8π}{3}G₄\left[Λ₄+\frac{1}{2π²a⁴}\left(\frac{M₊}{k₊}-\frac{M_-}{k_-}\right)\right].
    \end{equation}
    While this can be used to explain an epoch of radiation domination, we also see that a bulk particle that `rides' the bubble will also contribute as a radiation density.  One way to understand this is that the effect of climbing the out of the gravitational well associated to the AdS throat results in an additional gravitational redshift to the mass of particles.
    Drawing inspiration from the Friedmann equation above, we see that in order to have four-dimensional matter which redshifts as $1/a³$, one needs five-dimensional matter with a mass $M\sim r$, \ie, an object whose mass increases with $r$. 
    A string with a constant tension could be one such object whose mass would go as $M\sim α r$, where $α$ is the tension of the string. 

In particular, we will look for a four-dimensional Friedmann equation with matter in the form of homogeneous dust. We start out with a five-dimensional asymptotically AdS black hole solution sourced by  a uniform distribution of strings extending along the radial direction. The metric  was constructed forty years ago and goes by the name of `cloud of strings' \cite{Letelier:1979ej, Stachel:1980zr}. We briefly review the construction of this solution in AdS₅ as presented in \cite{Chakrabortty:2011sp, Dey:2017xty}.
The Einstein's equations in the presence of a negative cosmological constant $\tilde{\Lambda}_5$ are
\begin{equation}
R_{\mu\nu} - \frac{1}{2} R g_{\mu\nu} + \tilde{\Lambda}_5 g_{\mu\nu} = 8 \pi G_5 T_{\mu\nu},
\label{e-e}
\end{equation}
where $T_{\mu\nu}$ is the stress-energy tensor corresponding to a uniform distribution of strings,
\begin{equation}
\label{SE}
T^{\mu\nu} = - T \sum_i \int d^2\xi \frac{1}{\sqrt{|g_{\mu\nu}|}} \sqrt{|\sigma_{\alpha\beta}|} {{\sigma^{\alpha\beta}}} \partial_\alpha X^\mu 
\partial_\beta X^\nu \delta_i^5 (X - x_i).
\end{equation}
As before, $g_{\mu\nu}$ is the five dimensional
metric, while $\sigma_{\alpha\beta}$ now represents the induced metrics on the string worldsheets parametrized by $\xi$ with tension $T$. The delta functions appear due to the presence of localized strings, with $X$ labeling bulk coordinates and $x_i$ referring to the position of the $i^{\rm th}$ string. Two dimensions of the delta function correspond to the reparametrization freedom of the string worldsheet and are integrated out after fixing a particular gauge, for instance the static gauge, $t = \xi^0$ and $r = \xi^1$.

The smeared stress tensor $\mathcal{T}_{μν}$ can be defined as the integral of the stress tensor over a sphere. This gives
\begin{equation}
	\mathcal{T}_μ^ν = \int T_μ^ν \, d^3\theta = -\frac{3}{8 \pi }\frac{\alpha}{r^3} \delta_\mu^\nu,
\end{equation}
where $α$ is the average tension density of the strings defined as 
\begin{equation} 
\label{alpha}
\alpha = \frac{8 \pi}{3} {T \over V_3} \sum_{i=1}^{N} \int \delta_i^{(3)}(X - x_i)
\, d^3\theta
= \frac{8 \pi}{3} {T \over V_3} \sum_{i=1}^N 1 = {8 \pi \over 3} {TN \over V_3}.
\end{equation}
Here $N$ is the number of strings and $V_3$ is a dimensionless volume factor measuring the volume of the three-sphere at some reference time, $\tau_0$ defined by $a(\tau_0)=1$, (note that in our conventions the scale factor carries units of length).
Einstein's equations can be solved with this smeared stress tensor to give
\begin{equation}
\label{metric}
ds^2 = - f(r) dt^2 +
\frac{1}{f(r)} dr^2 + r^2 d\Omega_3^2
\end{equation}
with
\begin{equation}\label{eq:gttfull}
f(r) = 1 + k^2r^2 - \frac{8G_5M}{3\pi r^2} - \frac{2 G_5 \alpha}{ r}.
\end{equation}
Using this metric in the bulk, the Friedmann equation becomes
\begin{equation}\label{eq:frw_full}
    \frac{\dot{a}²}{a²}=-\frac{1}{a²}+\frac{8π}{3}G₄\left[Λ₄+\frac{1}{2π²a⁴}\left(\frac{M₊}{k₊}-\frac{M_-}{k_-}\right) 
    + \frac{3}{8πa³}\left(\frac{\alpha₊}{k₊}-\frac{\alpha_-}{k_-}\right)\right].
\end{equation}
As expected, the cloud of strings in five dimensions induces massive matter on the four-dimensional universe, with the endpoints of the five-dimensional strings representing massive particles in four dimensions.

It is straightforward to calculate the effective mass of a particle as seen from the four-dimensional braneworld in terms of tension of the string. 
As discussed before, the contribution of the mass term in $g_{tt}$ of the bulk metric given in \eqref{eq:gttfull} gives the radiation term in the effective Friedmann equation and one can read off its contribution to the energy density as 
\begin{equation}
\label{rho-scaling}
\rho_{\rm rad} \sim \frac{1}{2 \pi^2} \frac{M}{k} \frac{1}{a^4},
\end{equation}
where we take $M_+=M_-=M$ and the approximation comes from the assumption $k_+\sim k_-\sim k$. In order to extract the matter contribution to the Friedmann universe due to the string sources one can likewise read off
\begin{equation}
\label{rho-scaling-withT}
\rho_{\rm mat} \sim {3\over 8\pi}{\alpha \over k} {1\over a^3} = \frac{TN}{V_3k} \frac{1}{a^3}
\end{equation}
where $\alpha$ is given in \eqref{alpha}. From this scaling behavior of the energy density one can conclude that a string with constant tension $T$ appears as a particle with effective mass $m_{\text{eff}} = T/k$ residing on our braneworld.

To find phenomenologically low mass we would like to increase $k$ as much as possible. For gravity to be weakly coupled so that the geometrical picture makes sense, we need to have $k \ll M_5 \ll M_4$. A minimal hierarchy puts $M_5 \sim 10^{-1} M_4$ and $k \sim 10^{-3} M_4$. Furthermore, the tension of the strings cannot be lower than about $(\mathrm{TeV})^2$ without coming into conflict with collider data. From this we find a lower bound on particle masses given by
    \be
    m \sim \frac{(\mathrm{TeV})^2}{10^{-3} \times 10^{19} \mathrm{GeV}} \sim 10^{-1} \mathrm{eV}  .
    \ee
We find it intriguing that this is of the same order of magnitude as the estimated mass of the lightest known particles: the neutrinos. This is an encouraging and non-trivial observation. The low tension excludes fundamental strings and suggests the necessity of gauge strings or topological defects in the five-dimensional theory. To derive four-dimensional effective field theories matching the standard model will be an interesting, but challenging, problem.

\section{Gravity on the braneworld} \label{4dEinstein}

    Let us now examine the nature of four-dimensional gravity on the braneworld. This can be studied by considering the projection of the five-dimensional Einstein equations onto the brane using the Gauss equation,
    \begin{equation}\label{eq:Gauss}
    R^{(5)}_{αβγδ} e^α_c e^β_a e^γ_d e^δ_b = R^{(4)}_{cadb} + (K_{ad}K_{cb}-K_{cd}K_{ab}).
    \end{equation}
    Taking a trace yields
    \begin{equation} \label{eq:TrGauss}
    R^{(5)}_{αβγδ} e^α_c e^β_a e^γ_d e^δ_b h^{cd} = R^{(4)}_{ab} + (K_{ac}K^c_{b}-K^c_c K_{ab}) = \mathcal{J}_{ab}.
    \end{equation}
    Using contractions of the Gauss equation \eqref{eq:Gauss}, $\mathcal{J}_{ab}$ can be written as
    \begin{equation}\label{eq:curlyT}
    \begin{split}
 \mathcal{J}_{ab} &= R^{(5)}_{αβγδ} e^α_c e^β_a e^γ_d e^δ_b h^{cd}\\
    &= e^β_ae^δ_b\left(R^{(5)}_{αβγδ} e^α_c e^γ_d h^{cd}\right)\\
    &= e^β_ae^δ_b \left(R^{(5)}_{βδ} - R_{μβνδ}n^μ n^ν\right).
    \end{split}
    \end{equation}
    Assuming that the extrinsic curvature of the brane is dominated by the five-dimensional cosmological constant, we can write $K_{ab} = k h_{ab} + τ_{ab}$, where $h_{ab}$ is the induced metric on the brane and
    $τ_{ab}$ is sub-leading compared to $k$.
    Using this approximation \eqref{eq:TrGauss} becomes
    \begin{equation}
     \mathcal{J}_{ab}-R^{(4)}_{ab} = -3k²h_{ab}-k\left(2τ_{ab}+τh_{ab}\right) + \mathcal{O}\left(\frac{τ²}{k²}\right),
    \end{equation}
   where a tensor written without indices represents the trace taken with respect to the induced metric, $h_{ab}$. Taking the difference of this equation across the brane gives (ignoring terms of order $τ²$ and higher)
    \begin{equation}\label{eq:diffcurlyT}
    \left(\frac{ \mathcal{J}^{-}_{ab}}{k_-} - \frac{ \mathcal{J}^{+}_{ab}}{k_+}\right)
    - R^{(4)}_{ab}\left( \frac{1}{k_-} - \frac{1}{k_+} \right)
    = 3\left(k_+ - k_-\right)h_{ab} + 2\left(τ^+_{ab}-τ^-_{ab}\right) + \left(τ^+-τ^-\right)h_{ab}+ \mathcal{O}\left(\frac{τ²}{k²}\right).
    \end{equation}
    Defining $t_{ab}=-\left(K^+_{ab}-K^-_{ab}\right)/\left(8πG₅\right)$, the junction condition \eqref{eq:israel} can be written as
    \begin{equation}
    S_{ab}= -\frac{1}{8πG₅}\left[\left(K^+_{ab} - K^-_{ab}\right) - \left(K^+ - K^-\right)h_{ab}\right]\equiv t_{ab} - th_{ab},
    \end{equation}
    and \eqref{eq:diffcurlyT} can be written in terms of $t_{ab}$ as
    \begin{equation}
    \left(\frac{ \mathcal{J}^{-}_{ab}}{k_-} - \frac{ \mathcal{J}^{+}_{ab}}{k_+}\right)
    - R^{(4)}_{ab}\left( \frac{1}{k_-} - \frac{1}{k_+} \right)
    = h_{ab}\left[3\left(k_--k_+\right)-8πG₅t\right]-16πG₅t_{ab} +\mathcal{O}\left(\frac{τ²}{k²}\right),
    \end{equation}
    which can be rearranged to give
    \begin{equation}
    R^{(4)}_{ab} = \left(\frac{k_+k_-}{k_--k_+}\right)\left[h_{ab}\left(3\left(k_--k_+\right) - 8πG_5t\right)-16πG₅t_{ab} + \left(\frac{\mathcal{J}^{+}_{ab}}{k_+} - \frac{\mathcal{J}^{-}_{ab}}{k_-}\right) \right].
    \end{equation}
    This can be trace reversed to compute the four-dimensional Einstein tensor $G^{(4)}_{ab} 
    = R^{(4)}_{ab} - (R/2)g_{ab}$, \ie,
    \begin{equation}
    G^{(4)}_{ab} = 
    \left(\frac{k_+k_-}{k_- - k_+}\right) \bigg[ \left(\frac{\mathcal{J}^{+}_{ab}}{k_+} - \frac{\mathcal{J}^{-}_{ab}}{k_-}\right) - \frac{1}{2}h_{ab}\left(\frac{\mathcal{J}^{+}}{k_+} - \frac{\mathcal{J}^{-}}{k_-}\right)
    - 3(k_- -k_+)h_{ab} -16πG_5(t_{ab}-th_{ab})\bigg].
    \end{equation}
    This equation allows us to see that in addition to expected contributions to the Einstein equations coming from the stress tensor on the brane via the last terms in the square brackets, there are also contributions from five-dimensional geometry via the $\mathcal{J}$ tensor.
    
    The stress tensor of an empty brane with tension $σ$ is given by $S_{ab}=-σh_{ab}$, which in terms of $t_{ab}$ gives $t_{ab}=(σ/3)h_{ab}$ and $t=(4/3)σ$. In terms of $σ$, the $G^{(4)}_{ab}$ can be written as 
    \begin{equation}\label{eq:4dEinstein}
    G^{(4)}_{ab} = h_{ab}\left[16πG₅σ\left(\frac{k_+k_-}{k_- - k_+}\right)-3k_+k_-\right]
    + \left(\frac{k_+k_-}{k_- - k_+}\right) \left[ \left(\frac{\mathcal{J}^{+}_{ab}}{k_+} - \frac{\mathcal{J}^{-}_{ab}}{k_-}\right) - \frac{1}{2}h_{ab}\left(\frac{\mathcal{J}^{+}}{k_+} - \frac{\mathcal{J}^{-}}{k_-}\right) \right].
    \end{equation}
    To investigate the nature of the four-dimensional Einstein equations, let us pick the bulk metric. As discussed in the previous section, we consider the bulk spacetime to be asymptotically AdS₅ with a uniformly dense cloud of strings. This corresponds to \eqref{eq:gttfull},
    \begin{equation}
    f(r) = 1 + k²r²- \frac{8G₅M}{3πr²} - \frac{2G₅\alpha}{r}.
    \end{equation}
    Placing the brane at $r=a(τ)$ gives
    \begin{equation}
    \frac{\mathcal{J}_{ab}}{k}= h_{ab}\left(-3k+\frac{8G₅M}{3πka(τ)⁴}+\frac{3G₅\alpha}{ka(τ)³}\right) - \left(\frac{32G₅M}{3πka(τ)⁴}+\frac{6G₅\alpha}{ka(τ)³}\right)δ_a^0 δ^0_b,
    \end{equation}
    which when inserted into \eqref{eq:4dEinstein} gives
    \begin{equation}\label{eq:T4}
    \begin{split}
\left(G^{(4)}\right)^a_{b} =&- \underbrace{2k_+ k_-\left(3 -\frac{8πG₅}{k_--k_+}σ\right)}_{\equiv 8πG_4\left(σ_\textrm{crit}-σ\right) \equiv  8πG_4Λ_4}
        δ^a_b 
-\frac{8G₅}{πa(τ)⁴}\left(\frac{M_+k_- - M_-k_+}{k_--k_+}\right)
        \left(δ^a_0δ^0_b-\frac{1}{3}\sum_{i=1}^3δ^a_iδ^i_b\right)\\
        &-\frac{6G₅}{a(τ)³}\left(\frac{\alpha_+k_- - \alpha_-k_+}{k_--k_+}\right)δ^a_0δ^0_b
        +16πG_5\left(\frac{k_+ k_-}{k_- - k_+}\right)\left(T_\textrm{brane}\right)^a_b,
    \end{split}
    \end{equation}
    where the right hand side has been split up into three pieces corresponding to a four-dimensional cosmological constant ( $p=-ρ$), four-dimensional radiation ($p=ρ/3$) and pressure-less dust ($p=0$).
    The last term $T_\textrm{brane}$ corresponds to the stress tensor coming from worldvolume matter on the brane. We have used the definitions \eqref{eq:G4L4} to identify the term on the left hand side of the above equation with the four-dimensional cosmological constant.

    Since the right hand side of \eqref{eq:T4} has three pieces which go as the cosmological constant, radiation and matter, it is clear that the evolution of the brane is given by the Friedmann equation with these components. We can see this explicitly by computing the $(τ,τ)$ component of the induced Einstein tensor computed from the FLRW metric on the brane, 
    \begin{equation}
    G^{(4)}_{ττ} = 3\left(\frac{1}{a(τ)²} + \frac{a^\prime(τ)²}{a(τ)²}\right).
    \end{equation}
    Comparing this with \eqref{eq:T4} gives 
    \begin{equation}
    \begin{split}
    \frac{\dot{a}²}{a²}=
    -\frac{1}{a²}+\frac{8πG_4}{3}Λ₄ 
    -\frac{8πG_4}{3}\left(T_\textrm{brane}\right)_{00}
    +\frac{4G_4}{3πa^4}\left(\frac{M_+}{k_+}-\frac{M_-}{k_-}\right)+\frac{G_4}{a^3}\left(\frac{\alpha_+}{k_+}-\frac{\alpha_-}{k_-}\right),
    \end{split}
    \end{equation}
    where, like in \eqref{eq:T4}, we have shown the contribution of matter fields on the brane explicitly.
It should be noted that the energy momentum on the brane contributes with a {\it negative} sign to the net energy momentum in the Friedmann equations.  A related phenomenon concerns the negative contribution from the interior metric through $M_-$. 

It is important to understand that the actual energy that a four-dimensional observer measures is the combination of the energy on the brane and the effect of five-dimensional geometry.  How worldvolume fields couple to and backreact on geometry will depend on the precise brane embedding and the theory under consideration.  Without these couplings in hand, it is difficult to predict the nature of gravitational interactions of worldvolume fields.  For instance, if the presence of worldvolume fields backreacts onto the bulk metric, then all processes taking place in four dimensions can be thought of as low energy shadows of processes that could be taking place at very high energies in five dimensions, where no troublesome signs arise.  Particularly, the negative sign in front of $T_{({\rm brane})}$ may evoke fears of tachyons, however as long as the worldvolume fields inherited from string theory are not tachyonic, there can be no catastrophes at high energy.  While low dimensional couplings may look exotic, one would need a model specific set of field theoretic couplings to fully understand four-dimensional phenomenology.  This is obviously a challenging direction for future work. 

\section{Discussion and outlook} \label{outlook}

Our model suggests that dark energy can be incorporated into string theory if we allow for time dependence in an extra dimension. Rather than insisting on a time independent vacuum, we make use of a phase transition which places our universe on an expanding bubble. There are several interesting and important questions to explore.

To make sure that our model can be embedded into string theory, we must find an explicit ten-dimensional vacuum with the desired properties. We took the first few steps in  \cite{Banerjee:2018qey}. In particular, we need a sufficiently rich landscape of non-supersymmetric AdS$_5$ vacua so that we can find a small enough cosmological constant compatible with observations.  In addition to finding a phenomenologically viable model in string theory, any explicit embedding of this process will be useful in answering open questions in string theory.  In addition to possibly disproving the dS swampland conjecture, understanding the decay of AdS in detail may provide answers to questions about the validity of non-supersymmetric holography raised by \cite{Ooguri:2016pdq,Freivogel:2016qwc} and the proposed counter-example \cite{Giombi:2017mxl}. 

A proper holographic interpretation of our model essentially involves three parts. First, an understanding of the nucleation process from the perspective of the dual field theory living on the boundary of AdS, second, a CFT description of the expansion of the bubble universe and the energy cost thereof and third, the effective field theory on the bubble universe itself. In \cite{Barbon:2010gn}, a holographic field theory dual to Coleman-de Luccia \cite{Coleman:1980aw} bubble was proposed in terms of an instanton action in the scalar sector of an unstable deformed CFT. This duality was established by matching the decay rates in the bulk and the boundary instanton actions. Once the bubble nucleates, the subsequent expansion can be interpreted as a renormalization group (RG) flow mediated by a relevant or marginally relevant deformation of the boundary CFT. This notion was cultivated recently in a simpler setting of AdS$_3$/CFT$_2$ correspondence \cite{Antonelli:2018qwz} where the afore-mentioned connection between the expansion of the bubble and the RG flow in the boundary CFT was argued in terms of the flow of holographic entanglement entropy along the expansion, although the exact nature of the relevant deformation responsible for the RG flow was not known from this analysis. One might expect the expansion to be dual to a double trace deformation of the boundary CFT similar in spirit of \cite{Bernamonti:2009dp}. Other related studies of holographic braneworlds and decaying AdS cosmologies in the context of holography have been undertaken in  \cite{ Hawking:2000kj, Hertog:2004rz, Hertog:2005hu}. However, it would be interesting to investigate all these ideas in our set-up, particularly in presence of  string sources. 
 
 In an asymptotically AdS spacetime, strings stretching in the radial direction is the natural way to get massive particles in holography \cite{Herzog:2006gh, Liu:2006ug, Gubser:2006bz}. The presence of non-normalizable modes in these string backgrounds can change the structure of the boundary RG flow implied by an expanding brane. In particular, it would be worth investigating if in presence of these modes, we can set up a holographic dual to the nucleation and eventual expansion in terms of a full time-dependent RG flow.\footnote{One should not think of the dual set up in \cite{Barbon:2010gn} as a time-dependent RG flow since the Fubini instanton are normalizable configurations on the boundary CFT. We are grateful to Jose L.F. Barbon for pointing this out.} Another interesting approach to understand the holography of our model would be to study Green's functions both on the boundary CFT and for the effective field theory on the expanding brane. Similar computations  for collapsing shells were done in \cite{Danielsson:1998wt, Danielsson:1999zt, Danielsson:1999fa}, where the motion of the shell in the bulk was shown to correspond to unstable excitations appearing as poles in the boundary Green's function. This computation can be generalized for an expanding shell and in the presence of string sources. Furthermore, in this line of approach, one can also compute the Green's function on the expanding shell providing with an independent way to understand effective gravitational field theory on the braneworld. Last, but not  least, it would be interesting to connect the Green's function on the expanding shell to that on the AdS boundary in the light of recently proposed $T\overline{T}$ deformation of boundary CFT \cite{McGough:2016lol, Taylor:2018xcy}. This would be instrumental in understanding apparent (non) localization of gravity on our time-evolving braneworld. We shall report on progress in this direction  soon.

We also need to address the early universe. Can inflation be realized in our model?  While inflation is a quasi-dS phase, an exact dS period of inflation is now completely ruled out by $n_s = 0.968 \pm 0.006$ \cite{Ade:2015xua}.  Therefore, additional ingredients will be necessary to alter the spectral tilt and find an instance of slow-roll inflation consistent with observations.  Interestingly, the eventuality of bubble collisions in our model provides potentially promising pathways for inflationary model building. It was shown in \cite{DAmico:2014ywj} that brane collisions can result in free passage for the branes for a certain range of relativistic scattering velocity.  Additionally, open strings stretching between the branes will be created in the collision.  The model \cite{DAmico:2012wal} makes use of repeated collisions of this sort to provide an explanation for certain anomalies in the CMB power spectrum \cite{DAmico:2013hur}.  Finally, due to the creation of massive particles, collisions potentially provide a transition from a cosmological constant to a matter domination era.

Matter interactions and the realization of the standard model of particle physics is another important issue. The presence of branes and stretched strings suggest that the familiar D-brane constructions of gauge theories will be relevant. An important new feature is that it is now obvious that we need more than just fundamental strings. 
The endpoint of a string of tension $T$ appears as a particle of mass $T/k$ in the four-dimensional shellworld. Assuming a large five-dimensional cosmological constant (with no hierarchy across the shell) and setting a lower limit on the tension of strings at (TeV)², sets a lower limit on the particle mass to be of the order of magnitude of $0.1$eV, which is, interestingly enough, of the same order of magnitude as the estimated mass of neutrinos. This is an encouraging and non-trivial observation. The low tension excludes fundamental strings and suggests that the five-dimensional stretched strings need to be either topological defects or gauge strings. To derive four-dimensional effective field theories matching the standard model will be an interesting, but challenging, problem.

Finally, the incorporation of black holes into our model is an interesting direction for future research. The formation of a four-dimensional black hole is expected to proceed through the collision of stretched strings in five dimensions. Hence, one would expect a black hole to correspond to a five-dimensional black string ending on the braneworld. A problem with such a solution is that it may suffer from a Gregory-Laflamme instability \cite{Gregory:1993vy}. 
In \cite{Danielsson:2017riq,Danielsson:2017pvl}, it was argued, in four dimensions, that black holes can be replaced by \emph{horizon-less black shells}. If there is an instability of Minkowski space towards the formation of a bubble of AdS-space, it can be argued that a transition can be stimulated to occur if matter threatens to form a black hole. In \cite{Danielsson:2017riq} it was also argued that such black shells can be stable and thus correspond to a viable alternative to a black hole. The uplift of such a transition, with the black string  replaced by a black tube, seems as a natural outcome in our model. Results of this approach will be reported elsewhere.

\section*{Acknowledgements}
We are very grateful to Jose L.F. Barbon, Ivano Basile, Adam Bzowzki, Ben Freivogel, Thomas Hertog, Samir Mathur, Miguel Montero, Sudipta Mukherji, Lisa Randall, Ashoke Sen, Savdeep Sethi, Gary Shiu, Paul Steinhardt, Jesse van Muiden and Thomas Van Riet for their comments and very useful and interesting discussions. This work was partially performed at the Aspen Center for Physics, which is supported by National Science Foundation grant PHY-1607611. GD would also like to thank the FISPAC (University of Murcia) for warm hospitality while part of this work was completed. The work of UD, GD, SG, and MS is supported by the Swedish Research Council (VR).
The work of SB is supported by the Knut and Alice Wallenberg Foundation under grant 113410212.

\bibliographystyle{JHEP}
\bibliography{bib}

\end{document}